\documentclass[journal]{IEEEtran}
\usepackage[latin1]{inputenc}
\usepackage{amsmath,amssymb,amscd,latexsym,dsfont }
\usepackage{algpseudocode}
\usepackage{algorithm}
\usepackage{algorithmicx}
\usepackage{comment,psfrag,enumerate}
\usepackage[caption=false, font=footnotesize, aboveskip=0pt]{subfig}
\usepackage{cite}
\usepackage{url}
\usepackage{tabularx}
\usepackage{graphicx}
\usepackage{pstricks}
\usepackage{float,color}
\usepackage{rotating}
\usepackage{multirow}
\usepackage{multicol}
\usepackage{stfloats}
\usepackage{setspace}
\newcommand{\eqlab}[2]{\begin{align} \label{#1} #2 \end{align}}
\newcommand{\eq}[1]{\begin{align*} #1 \end{align*}}
\newcommand{\eqrf}[1]{\eqref{#1}}
\newtheorem{theo}{Theorem}
\newtheorem{lema}{Lemma}
\usepackage{hyperref}

\definecolor{DarkGreen}{rgb}{0.0, 0.5, 0.0}

\begin{document}

\captionsetup[subfloat]{farskip=0pt}

\title{A Low-Complexity Detector for Memoryless
Polarization-Multiplexed Fiber-Optical Channels}

\author{Authorlist to be added \thanks{Affiliations to be added}}

\author{Christian~H\"ager, \emph{Student Member,
IEEE}, Lotfollah~Beygi, \emph{Member, IEEE},  Erik~Agrell, \emph{Senior
Member, IEEE},
Pontus~Johannisson, Magnus~Karlsson, \emph{Member, IEEE}, and Alexandre Graell i Amat,
\emph{Senior Member, IEEE}
\thanks{The authors are with Chalmers University of Technology, Sweden
(e-mail: \{christianhaeger, beygil, agrell, pontus.johannisson,
magnus.karlsson, alexandre.graell\}@chalmers.se). 
This work was partially funded by the Swedish Research Council
under grant \#2011-5961, by the Swedish Foundation for Strategic
Research (SSF) under grant no.~RE07-0026, and by the Swedish
Governmental Agency for Innovations Systems (VINNOVA/CELTIC) under
grant 2010-01238.}%
}

\maketitle


\begin{abstract}
A low-complexity detector is introduced for polarization-multiplexed
$M$-ary phase shift keying modulation in a fiber-optical channel
impaired by nonlinear phase noise, generalizing a previous result by
Lau and Kahn for single-polarization signals. The proposed detector
uses phase compensation based on both received signal amplitudes in
conjunction with simple straight-line rather than four-dimensional
maximum-likelihood decision boundaries.
\end{abstract}

\vspace{-0.5cm}

\section{Introduction}

%

The Manakov equation describes the propagation of a
polarization-multiplexed signal in a fiber-optical channel. Two major
impairments, linear chromatic dispersion and the Kerr nonlinear
effect, are modeled by this equation. The nonlinear effect causes a
phase rotation proportional to the field instantaneous power. The
interaction of the signal and the amplified spontaneous emission (ASE)
noise generated by optical amplifiers due to the nonlinear Kerr effect
gives rise to nonlinear phase noise (NLPN). NLPN imposes a major degradation
in the performance of coherent optical data transmission systems.

Bononi \emph{et al.} \cite{NLPN_Bononi_modulation_format} investigated
the effect of NLPN on popular modulation formats for single-channel
and wavelength-division multiplexing systems in a dispersion-managed
fiber link. The performance of orthogonal frequency-division
multiplexing systems in the presence of NLPN has been evaluated in
\cite{NLPN_OFDM} by theoretical, numerical, and experimental
approaches. In \cite{NLPN_Demir},\cite[ch. 4]{Keang_Po_HO_book},
comprehensive surveys of known techniques for the analysis and
characterization of NLPN and its impact on the system performance are
provided. 
 
The statistics of NLPN and the detector design for a channel with NLPN
have been studied in \cite{Mecozzi_04,NLPN_Kumar,NLPN_Yadin} by
analytical approaches and in \cite{Chritian_Hager2013} by numerical
methods. The joint probability density function (pdf) of the received
amplitude and phase given the initial amplitude and phase of the
transmitted signal and the optical signal-to-noise ratio (OSNR) is
derived in \cite{turitsyn_information_2003, Mecozzi_04,
yousefi_per-sample_2011} \cite[pp. 157, 224--225]{Keang_Po_HO_book}
for a fiber-optical channel with NLPN caused by distributed or lumped
amplification. 
Moreover, compensation of NLPN has been studied in
\cite{Lau_Kahn_NLPN_ML_compensator} based on the aforementioned pdf.

In this paper, we extend the detector structure introduced for a
single polarization (SP) $M$-PSK system in
\cite{Lau_Kahn_NLPN_ML_compensator} to polarization-multiplexed (PM)
$M$-PSK, using the signal statistics derived in
\cite{beygi_JLT_DM_DP}. 
To this end, we first introduce a simplified approach to reproduce the
result in \cite{Lau_Kahn_NLPN_ML_compensator} for the SP case. This
method can be easily used to extend the result to the PM case
and can also be applied to both lumped and distributed amplification.
For simplicity, we assume single-channel transmission and
inter-channel effects are not taken into consideration.
The symbol error rate (SER) of the proposed detector is compared to
the performance of the maximum-likelihood (ML) detector for PM-$4$-PSK
and to the performance of the ML detector for SP-$4$-PSK for the same
bandwidth as well as for the same data rate.



\section{System Model and Preliminaries} \label{SEC:Model}

We assume zero dispersion to make the analysis applicable to
memoryless (nondispersive) fiber-optical channels, similarly as in,
e.g.,
\cite{Gordon_mollenaur_90,Mecozzi_04,Chritian_Hager2013,yousefi_per-sample_2011,Lau_Kahn_NLPN_ML_compensator}.
Due to this assumption, the subsequent analysis ignores the
interaction of chromatic dispersion and nonlinearity. The resulting
model can serve as an
approximation for dispersion-managed transmission links provided that
the local accumulated dispersion is sufficiently low
\cite{NLPN_Bononi_modulation_format, beygi_JLT_DM_DP}. 
For a zero polarization-mode and chromatic dispersion fiber-optical
channel, the Manakov equation with loss included reduces to \cite[ch. 6]{Book_agrawal}
\eqlab{eq:NLSE}{
j{\partial \mathbf{E} \over \partial z} + \gamma(\mathbf{E}\mathbf{E}^{\dag})\mathbf{E}+j{\alpha \over 2}\mathbf{E}=\mathbf{0},
} 
where $\mathbf{E} = (E_\text{x},E_\text{y})$ is the polarization-multiplexed
launched envelope signal into the fiber, $\gamma$ is the nonlinear coefficient, $\alpha$ is the attenuation coefficient, $\dag$ denotes Hermitian conjugation, and $z$ is the distance from the beginning of the fiber. 
The solution to \eqrf{eq:NLSE} at time $t$ can be written as \cite[ch. 4]{Book_agrawal} \eqlab{eq:NLSE-solution}{
\mathbf{E}(z,t) = \mathbf{E}(0,t)q(z)\exp\left(j\gamma P_{0}(t)
\int_0^z q^2(\tau) \, \text{d} \tau \right),
}
where $P_{0}(t) = |E_\text{x}(0,t)|^{2}+|E_\text{y}(0,t)|^{2}$ is the
instantaneous launched power into the fiber and $q(z) = \exp(-\alpha
z/2)$ is a function that describes the power evolution. 

Here, we assume a fiber link with total length $L$ and either
distributed or lumped amplification to compensate for the fiber loss
perfectly. 
%
We consider ASE noise within the optical signal bandwidth, i.e.,
ignoring the Kerr effect induced from out-of-band signal and noise in
the same way as in \cite{Gordon_mollenaur_90}. 
If a four-dimensional (4D) signal $\mathbf{S} =
(S_\text{x},S_\text{y})$, consisting of two two-dimensional (2D)
complex signals, is transmitted, it experiences
an overall NLPN $\phi_\text{NL} = \phi_\text{x} + \phi_\text{y}$. The
terms $\phi_\text{x}$ and $\phi_\text{y}$ are generated by the
interaction of the signal and noise due to the Kerr effect in
polarizations x and y, respectively. 
For lumped amplification and a link consisting of $N$
spans, 
ASE noise $n_\text{x}^i$, $i=1, \ldots, N$, with variance $\sigma_0^2$
is added after each span.\footnote{Throughout the paper, we give expressions for 
polarization x only, if polarization y has an equivalent
expression.} 
One may use \eqrf{eq:NLSE-solution} to obtain  $\phi_\text{x} = \gamma
L_{\text{eff}}\sum_{i=1}^N |S_\text{x} +
\sum_{l=1}^{i}n_\text{x}^l|^{2}$, where $L_\text{eff} =
(1-\exp(-\alpha L/N))/\alpha$ is the effective nonlinear
length. It is clearly seen that signals in both
polarizations contribute to the generated NLPN $\phi_\text{NL}$.
The received electric field $\mathbf{E}$ can be written as
$\mathbf{E} = \hat{\mathbf{E}}e^{-j\phi_\text{NL}}$,
where $\hat{\mathbf{E}} = \mathbf{S} +  \sum_{i = 1}^{N} \mathbf{n}^i$ is the linear part of the electric field and $\mathbf{n}^{i} = (n_\text{x}^{i}, n_\text{y}^{i})$.
One may regard distributed amplification as lumped amplification with
an infinite number of spans. This gives $\lim_{N \rightarrow \infty}
NL_\text{eff}  = L$. 
In this case, a continuous amplifier noise vector~$\mathbf{n}(z)=
(n_\text{x}(z),n_\text{y}(z))$ is considered. The elements of this
vector are 
zero-mean complex-valued Wiener processes
\cite[p.~154]{Keang_Po_HO_book} with autocorrelation function  $
\text{E}[n_\text{x}(z_1)n_\text{x}^*(z_2)]\!\!=
\sigma_\text{d}^{2}\text{min}(z_1,z_2),$ where $\sigma_\text{d}^{2} = 2h\nu_\text{opt} W \alpha n_\text{sp}$
\cite{Lau_Kahn_NLPN_ML_compensator}, $h\nu_\text{opt}$ is the
energy of a photon, $n_\text{sp}$ is the spontaneous emission factor,
and $W$ is the bandwidth of the optical signal.
The SNR vector is defined as $\boldsymbol\rho = (\rho_\text{x} , \rho_\text{y})$ where $\rho_\text{x}$ is $|S_\text{x}|^2/(L\sigma_\text{d}^{2})$ or $|S_\text{x}|^2/(N\sigma_{0}^{2})$ for distributed or lumped amplification, respectively.
The normalized received amplitude $r_\text{x}$ is denoted by
$|E_\text{x}|/(\sigma_\text{d}\sqrt{L})$ or by
$|E_\text{x}|/(\sigma_0\sqrt{N})$ for distributed and lumped
amplifications. 

The joint pdf of the received phase vector
$\boldsymbol\theta=(\theta_\text{x},\theta_\text{y})$ and the
normalized amplitudes $\mathbf{r} = (r_\text{x},r_\text{y})$ of a
zero-dispersion fiber-optical channel is \cite{beygi_JLT_DM_DP}
\eqlab{eq:Joint_pdf_continous}{
&f_{\boldsymbol\Theta,\mathbf{R}}(\boldsymbol\theta,\mathbf{r}) ={f_\mathbf{R}(\mathbf{r}) \over 4\pi^{2}}  +{1 \over 2\pi^2}\sum\limits_{k_\text{x} = 1}^\infty \text{Re}\left\{C_{\mathbf{k}_\text{x}}(\mathbf{r})  e^{jk_\text{x}\theta_\text{x}} \right\} \nonumber\\
&\qquad + {1 \over 2\pi^{2}}\sum\limits_{k_{\text{x}} = 1}^\infty\sum\limits_{k_{\text{y}} = 1}^\infty \text{Re}\left\{ C_\mathbf{k}(\mathbf{r})  e^{j\mathbf{k}\cdot\boldsymbol\theta} + C_\mathbf{k^{*}}(\mathbf{r})  e^{j\mathbf{k^{*}}\cdot\boldsymbol\theta}\right\}\nonumber\\&\qquad\qquad\qquad+
{1 \over 2\pi^2}\sum\limits_{k_\text{y} = 1}^\infty \text{Re}\left\{C_{\mathbf{k}_\text{y}}(\mathbf{r})  e^{jk_\text{y}\theta_\text{y}}\right\},
}where $f_\mathbf{R}(\mathbf{r})$ is the joint pdf of the two normalized independent Ricean random variables $r_\text{x}$ and $r_\text{y}$, 
and the Fourier coefficients $C_\mathbf{k}(\mathbf{r})$ are given in
\cite{beygi_JLT_DM_DP}.
In \eqref{eq:Joint_pdf_continous}, we assume the transmitted phase
vector to be $(0,0)$. Due to the rotational invariance of the channel,
the pdf for an arbitrary transmitted phase vector $(\theta_{0,\text{x}},
\theta_{0,\text{y}})$ is obtained by replacing $\theta_\text{x}$ and
$\theta_\text{y}$ in \eqref{eq:Joint_pdf_continous} with
$\theta_\text{x} -
\theta_{0,\text{x}}$ and $\theta_\text{y} - \theta_{0,\text{y}}$, respectively.

For an SP scheme, the joint pdf of the phase and the normalized amplitude of the received signal in the corresponding polarization is simplified to \cite[ch. 5]{Keang_Po_HO_book}
\eqlab{eq:Joint_pdf_single}{
f_{\Theta,R}(\theta,r)\! =\! {f_R(r) \over 2\pi}\!+\!{1 \over
\pi}\sum\limits_{k = 1}^\infty \text{Re}\left\{C_k(r)
e^{j k \theta}\right\},}
where $f_{R}(r)$ is the Ricean pdf of the amplitude $r$, and the Fourier
coefficients $C_k(r)$ are given in \cite{beygi_JLT_DM_DP} for both
types of amplifications. 
Again, the transmitted phase in \eqref{eq:Joint_pdf_single} is assumed
to be 0, and the pdf for an arbitrary transmitted phase $\theta_0$ is
obtained by replacing $\theta$ with $\theta - \theta_0$. 

In the following, we consider $M$-PSK constellations with $s_k =
\sqrt{E_\text{s}} \exp\left( j{\pi \over M}(2k+1)\right)$, $k = 0,
\ldots, M-1$, where $E_\text{s}$ is the average energy of the
constellation. 

\section{The ML Receiver for SP-$M$-PSK} \label{SEC:3}

\begin{figure}[t]
	\centering
	\subfloat[]{\includegraphics{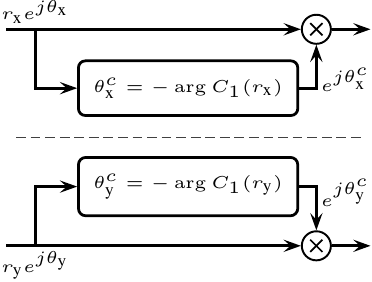}}
	\quad
	\subfloat[]{\includegraphics{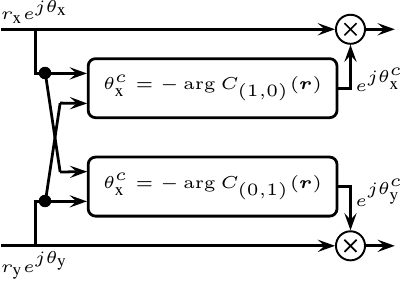}}
	\caption{Receiver for PM-$M$-PSK with (a) separate detection (PM-Det1) and (b) joint calculation of the amplitude-dependent phase rotations (PM-Det2).  
	}
	\label{fig:receiver}
	\vspace{-0.5cm}
\end{figure}

For SP-$M$-PSK, the optimal decision (Voronoi) regions for the
received constellation have spiral shape (cf. \cite[Fig.
1]{Lau_Kahn_NLPN_ML_compensator}), and hence ML detection is
computationally complex. To decrease the complexity of the detector,
Lau and Kahn showed that straight-line decision boundaries  can be
used, provided that an amplitude-dependent phase rotation $\theta^c$
is applied before detection \cite{Lau_Kahn_NLPN_ML_compensator}. The
corresponding receiver structure is illustrated in the top half of
Fig.~\ref{fig:receiver}(a). It can be seen that the phase rotation is
solely a function of the received amplitude in one polarization and a
simple ML detection of $M$-PSK for additive white Gaussian noise
(AWGN) with straight-line decision boundaries is subsequently
performed. 

In this section, we introduce a new approach to derive the optimal
phase rotation as a function of the received amplitude. In contrast to
\cite{Lau_Kahn_NLPN_ML_compensator}, this approach can be easily
extended to PM-$M$-PSK. For a transmitted phase of $\theta_0 = 0$, we
assume that the conditional pdf $f_{\Theta|R}(\theta|r)$ of the
received phase $\theta$ given the received amplitude $r$ is
approximately symmetric around $\theta_\text{max}(r)$, where
$\theta_\text{max}(r)$ denotes the phase value where
$f_{\Theta|R}(\theta|r)$ is maximum. This assumption is motivated by
inspection of the pdf and its validity is justified later by the
obtained results. In fact, an equivalent approximation was also done in
\cite[App.  A]{Lau_Kahn_NLPN_ML_compensator}. This assumption is used
for both distributed and lumped amplifications.

\begin{lema}\label{lemma_1}
	Let $f_{X}(x)$ be the (periodic) pdf of a random angle $X$.
	Furthermore, let the pdf be symmetric around $x_\text{max} \in
	(-\pi, \pi]$, the value where $f_{X}(x)$ has its maximum. If the pdf
	decreases monotonically from $x_\text{max}$ to $x_\text{max} \pm
	\pi$, then $x_\text{max} = -\arg\Psi_{X}(1)$, where $\Psi_{X}(\nu)$
	is the (discrete) characteristic function (CF) of $X$.

\end{lema}
\begin{IEEEproof}
	Define $\tilde{X} = X - x_\text{max}$. Since the pdf of $\tilde{X}$
	is an even function, its CF is real. Furthermore, the CFs of $X$ and
	$\tilde{X}$ are related via $\Psi_X(\nu) = \Psi_{\tilde{X}}(\nu)
	e^{-j \nu x_\text{max}}$. Letting $\nu = 1$ and solving for
	$x_\text{max}$ gives $x_\text{max} =
	\arg\Psi_{\tilde{X}}(1)-\arg\Psi_{X}(1)$. Thus, it needs to be shown
	that $\arg\Psi_{\tilde{X}}(1) = 0$. Having already established that
	$\Psi_{\tilde{X}}(1)$ is real, we only need to show that it is also
	positive. This follows from the definition $\Psi_{\tilde{X}}(1) =
	\int_{-\pi}^{\pi} f_{\tilde{X}}(x) \cos(x) \, \text{d} x$ and the
	fact that $f_{\tilde{X}}(x)$ is nonnegative and decreases monotonically
	from $0$ to $\pm \pi$. 
\end{IEEEproof}

Using Lemma 1, one can compute the rotation of the $M$-PSK ML decision
boundaries due to NLPN as described in the following theorem.

\begin{theo}\label{Theorem1}
Consider a memoryless fiber-optical channel with NLPN. 
The decision boundary of the ML detector for SP-$M$-PSK between
symbols $s_k$ and $s_{k+1}$ has the polar coordinates $(r,
\theta_b(k,r))$, where $\theta_b(k,r) =- C_1(r) + 2k/M$ for $r \ge 0$
and $C_1$ is the first Fourier coefficient in 
\eqref{eq:Joint_pdf_single}.
\end{theo}
\begin{IEEEproof}
The ML decision boundary between the two symbols of the $M$-PSK
constellation with $k = 0$ and $k = M-1$ is determined in such a way as to satisfy
\eq{
f_{\Theta|R,\Theta_0 }(\theta^\text{c}(r) \mid r,-\tfrac{\pi}{M})=
f_{\Theta|R,\Theta_0 }(\theta^\text{c}(r) \mid r,\tfrac{\pi}{M}).
}
Using the symmetry of
{$f_{\Theta|R,\Theta_0}(\theta|r,\theta_0)$} around
$\theta_\text{max}(r) + \theta_0$, we obtain
$\theta^\text{c}(r) = \theta_\text{max}(r)$. 
Using Lemma~1, we
get {$\theta_\text{max}(r) = -\arg~\Psi_{\Theta|R,\Theta_0}(1\mid
r,0) =  - \arg C_1(r)$}. 
\end{IEEEproof}



\section{Receivers for PM-$M$-PSK} \label{SEC:4}

For a fixed state of polarization, we receive two dependent 2D
symbols, which have been rotated by the NLPN equally. Using 
\eqref{eq:Joint_pdf_continous}, the ML detector in this case can
be written as 
\begin{align}
	{\boldsymbol{\hat{\theta}}}_0 = \arg\max_{\boldsymbol{\theta}_0}
	f_{\boldsymbol{\Theta}, \boldsymbol{R} | \boldsymbol{\Theta}_0}
	(\boldsymbol{\theta}, \boldsymbol{r} | \boldsymbol{\theta}_0).
	\label{eq:ML_detector}
\end{align}
The optimization is performed over all possible $M^2$
transmitted phase combinations for a PM-$M$-PSK signal. We refer to
this detector as ``PM-ML''. 

A simple, but clearly suboptimal, way to reduce the complexity of
solving \eqref{eq:ML_detector} is to {treat the received signals in
both polarizations independently. In other words, the marginal pdfs
$f_{\Theta_\text{x}, R_\text{x}}(\theta_\text{x}, r_\text{x})$ and
$f_{\Theta_\text{y}, R_\text{y}}(\theta_\text{y}, r_\text{y})$ are
used to perform detection separately in each polarization}, which
leads again to spiral-shaped decision boundaries as in Sec.
\ref{SEC:3}. Equivalently, one may extend the receiver structure for
SP in a straightforward manner as shown in Fig. \ref{fig:receiver}(a),
where a different rotation angle is applied to each received symbol,
based on the received amplitude in the corresponding polarization.
Using Theorem~1, the computation of the rotation angles is then based
on the first Fourier coefficient of the two marginal pdfs. We refer to
this detector as ``PM-Det1''. 

As seen in \eqrf{eq:NLSE-solution}, the phase rotation due to the
nonlinear Kerr effect is a function of the signal amplitudes in both
polarizations. Hence, one may improve the performance of PM-Det1 by
taking into account the amplitudes of both polarizations in computing
the phase rotation. To this end, we use the same
symmetry assumption as in
the previous section for $f_{\Theta_\text{x}|\mathbf{R}}(\theta|
\mathbf{r}) = f_{\Theta_\text{x},\mathbf{R}}(\theta,
\mathbf{r})/f_{\mathbf{R}}(\mathbf{r})$, where
$f_{\Theta_\text{x},\mathbf{R}}(\theta, \mathbf{r})$ is the marginal
of \eqrf{eq:Joint_pdf_continous} with respect to $\Theta_\text{y}$,
i.e., we assume that $f_{\Theta_\text{x}|\mathbf{R}}(\theta|
\mathbf{r})$ is symmetric around the phase for which this pdf is
maximum. 
This assumption allows us to describe the decision boundaries of the
PM-$M$-PSK signal distorted by NLPN in each polarization as the
rotated version of the straight-line decision boundaries for an AWGN
channel.
\begin{theo}\label{Theorem2}
The decision boundaries of the detector given by  
\eqlab{eq:DP_ML}{
	\hat{\theta}_{0,\text{x}} = \arg\max_{{\theta}_{0,\text{x}}}
	f_{\Theta_\text{x}, \boldsymbol{R} | {\theta}_{0,\text{x}}}
	({\theta}_\text{x}, \boldsymbol{r} | {\theta}_{0,\text{x}})
}
for polarization x can be transformed to straight lines using the phase rotation given by
\eqlab{eq:Rotation_DP}
{
\theta^\text{c}_\text{x}(\mathbf{r}) = - \arg C_{(1,0)}(\mathbf{r}),
}  
where $C_{(1,0)}(\mathbf{r})$ is the Fourier coefficient appearing in
\eqref{eq:Joint_pdf_continous} with $\boldsymbol{k}_x = (1,0)$.
Similarly, the rotation for polarization y is obtained as
$\theta^\text{c}_\text{y}(\mathbf{r}) = -\arg C_{(0,1)}(\mathbf{r})$. 
\end{theo}
\begin{IEEEproof}
One may follow an analogous approach as in the proof of
Theorem~\ref{Theorem1}, by replacing
$f_{\Theta_\text{x}|\mathbf{R}}(\theta| \mathbf{r})$ with
$f_{\Theta|R}(\theta|r)$ to show that the decision boundary between
symbols $S_x = s_{k}$ and $S_x = s_{k+1}$ in polarization x has
the parametric description
$r_\text{x}\exp(j\theta^\text{c}_\text{x}(\mathbf{r})+2jk\pi/M)$.
\end{IEEEproof}
The proposed detector implementing \eqref{eq:DP_ML} via this phase
rotation method is referred to as ``PM-Det2'' and shown in Fig.
\ref{fig:receiver}(b). It can be seen that the rotation angle in each
polarization is computed using the received amplitudes $\mathbf{r}$.
It is worth mentioning that since the rotation is an invertible
operation, joint 4D demodulation is still possible after the rotation.
For complexity reasons, however, we perform hard decision on each 2D
soft symbol using simple straight-line decision boundaries as shown in
the figure.

\begin{figure}[t]
	\begin{center}
		\subfloat[no comp.]{
				\includegraphics[width=2.2cm]{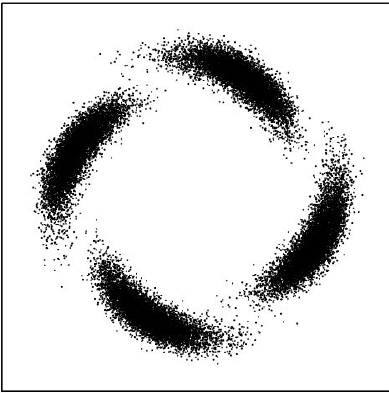}

		\label{fig:scatter_channel}
		}
		\subfloat[$\theta_{\text{x}}^c = f(r_\text{x})$]{
				\includegraphics[width=2.2cm]{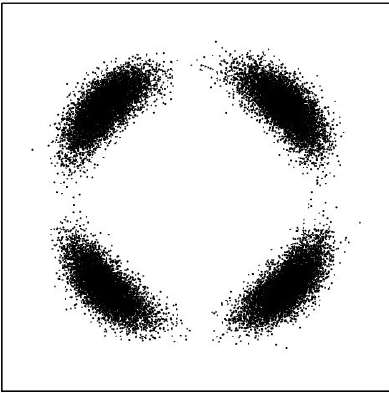}
		\label{fig:scatter_theta1}
		}
		\subfloat[$\theta_{\text{x}}^c = f(\boldsymbol{r})$]{
				\includegraphics[width=2.2cm]{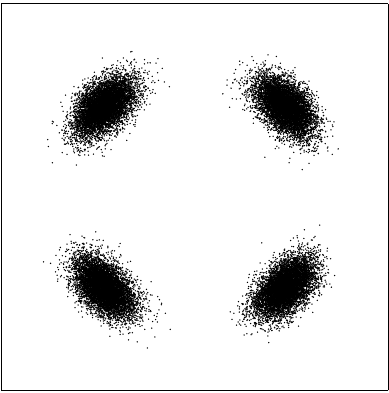}

		\label{fig:scatter_theta2}
		}
	\end{center}
	\caption{Scatter plots in the x polarization for (a) no compensation,
	(b) compensation according to Fig. 1(a), and (c) according to
	Fig. 1(b). Decision boundaries in (b) and (c) are
	straight lines. In (a), the boundaries are spiral shaped and
	depend on the received amplitude in the y polarization.} 
	\label{fig:regions}
\end{figure}

In Fig.~\ref{fig:regions}, a qualitative comparison of the two
different rotation schemes corresponding to PM-Det1 and PM-Det2 is
shown.  Fig.~\ref{fig:regions}(a) shows a scatter plot of the received
symbols in polarization x directly after the channel, i.e., no phase
compensation is assumed. In Fig.~\ref{fig:regions}(b), a phase
rotation of each received symbol is applied, which is solely based on
the corresponding received amplitude of this symbol (PM-Det1). Lastly,
Fig.~\ref{fig:regions}(c) shows the result of applying a phase
rotation that is based on the received amplitude of the received
symbols in \emph{both} polarizations (PM-Det2). Observe that the
second rotation method leads to a notably smaller phase variance
compared to the first method. However, it should be mentioned that the
receiver structure shown in Fig.~\ref{fig:receiver}(b) does \emph{not}
correspond to the ML receiver for PM-$M$-PSK, since the residual
phases in both polarizations after the rotation are not statistically
independent. The performance loss compared to ML detection is
quantified in the next section.

\section{Performance Analysis} \label{SEC:6}


The SER of PM-$M$-PSK for PM-Det1 and PM-Det2 can be computed
analytically. After the introduced phase rotations, the marginal pdf
of the phase in polarization x, given that the phase of the
transmitted signal is zero, is obtained by replacing $\theta_\text{x}$
with $\theta'_\text{x} - \theta^\text{c}_\text{x}$ in
$f_{\Theta_\text{x}|\mathbf{R}}(\theta| \mathbf{r})$ and then
integrating out the radii $r_\text{x}$ and $r_\text{y}$ over
$[0,+\infty)$ to get \eqlab{eq:Joint_pdf_x}{
f_{\Theta'_\text{x}}(\theta) = {1 \over 2\pi} +{1 \over
\pi}\sum\limits_{k= 1}^\infty
\cos(k\theta)\int\limits_{0}^{\infty}\int\limits_{0}^{\infty}
|C_{(k,0)}(\mathbf{r})|\text{d}r_\text{x}\text{d}r_\text{y}.} 
Here, we only show how to compute the SER of PM-Det2 for
a PM-$M$-PSK system. An analogous derivation can be applied for
PM-Det1. One can write
\eq{
&\text{SER}_\text{x} 
= 1-\int_{-{\pi \over M}}^{{\pi \over M}}f_{\Theta'_\text{x}}(\theta)\text{d}\theta
=
{M-1\over M}-
\nonumber\\
&\sum_{k=1}^{\infty}\frac{2\text{sinc}\left({k\over
M}\right)}{M}\int_{0}^{\infty}\!\!|C_{(k,0)}^\text{x}(r_\text{x})|\text{d}r_\text{x}\!
\int_{0}^{\infty}\!\!|C_{(k,0)}^\text{y}(r_\text{y})|\text{d}r_\text{y},
}
where $C_{(k,0)}^\text{x}$ and $C^\text{y}_{(k,0)}$ are computed
using \cite[eq. (26)]{beygi_JLT_DM_DP}.

\captionsetup[subfigure]{hangindent=5pt, indention=7pt, margin=15pt,
singlelinecheck=false, justification=Centering}

\begin{figure}[!t]
	\centering
	\subfloat[analytical SER with zero dispersion]{ \includegraphics[]{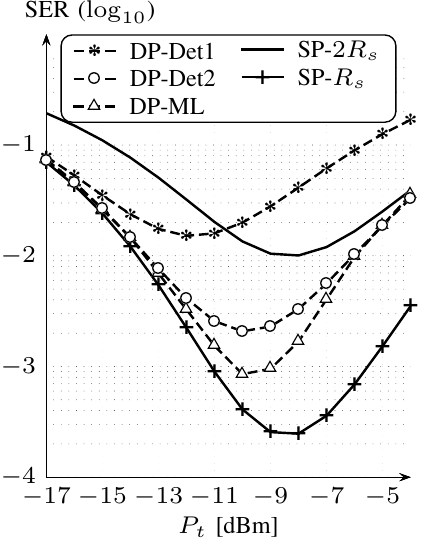}}
	\subfloat[simulated SER for a dispersion-managed link]{ \includegraphics[]{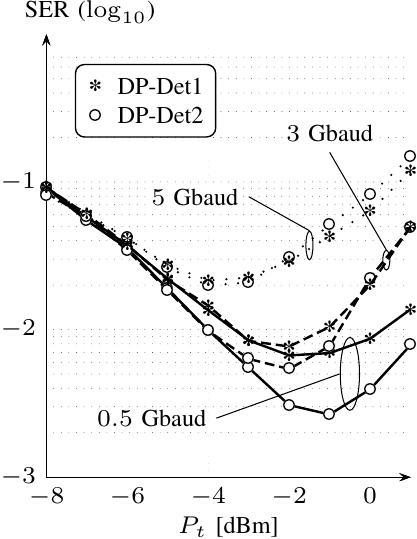}}

\caption{The SER of SP and PM systems with 4-PSK versus the
transmitted power per polarization $P_\text{t}$.  }
\label{fig:SER}
\vspace{-0.5cm}
\end{figure}

In Fig.~\ref{fig:SER}(a), the performance of PM-$4$-PSK is evaluated
using the above analytical approach for PM-Det1 and PM-Det2.  The SER
of the ML detector defined by \eqrf{eq:DP_ML} is given by a
four-dimensional integral of the pdf over the ML decision regions.
This SER, estimated by Monte-Carlo integration, is also shown in
Fig.~\ref{fig:SER}(a).  Moreover, we compute the SER of SP-$4$-PSK to
compare with SP data transmission in two different scenarios: (i) For
the same data rate per polarization (i.e., the same bandwidth) and
(ii) for the same total data rate as the PM case. This evaluation is
done for distributed amplification with channel parameters $L=9000$
km, $\gamma = 1.4~\text{W}^{-1}\text{km}^{-1}$, $R_\text{s}= 28$
Gbaud, $\nu_\text{opt} = 193.55$ THz, $\alpha = 0.25$ dB/km, and
$F_n=6$ dB. As seen in Fig.~\ref{fig:SER}(a), the PM schemes show
negligible performance degradation in the linear regime for a fixed
bandwidth (case (i)), i.e., for $P_\text{t}<-15$~dBm, compared to the
SP scheme.  For a fixed data rate (case (ii)), one may observe a $2$
dB performance improvement using PM-Det2, at a SER of
$1.5\times10^{-2}$. In the strongly nonlinear regime, the SP scheme is
superior to PM at the expense of losing half of the data rate.
Furthermore, in the linear regime, the SP scheme in case (i) and the
PM scheme have the same performance and their SER curves overlap,
while in the strongly nonlinear regime, the SER of the PM scheme
converges to the SER of the SP scheme in case (ii). This is because
the system performance is intimately related to the product of the
noise variance and the transmit power in the nonlinear regime, which
is the same for these two scenarios. Fig.~\ref{fig:SER}(a) also
indicates that in the linear regime, the detectors PM-Det2 and PM-Det1
perform similarly. However, the reduction in circular variance
observed in Fig.~\ref{fig:regions} translates into a noticeably better
SER in the nonlinear regime for PM-Det2 when compared to PM-Det1. In
the region of interest, i.e., SNRs around $-10$ dBm, the performance
degradation of PM-Det2 compared to the ML detector is $0.7$ dB. This
is due to independent detection of phase information in the two
polarizations. In Fig.~\ref{fig:SER}(b), we also show the performance
of DP-Det1 and DP-Det2 for a dispersion-managed link using the split
step Fourier method. The system parameters are the same as before, but
now we assume $45$ fiber spans of length $90$ km and a lumped
amplification scheme. Dispersion is compensated after each span using
an ideal dispersion-compensating fiber. The symbol rate is varied
between $0.5$ and $5$ Gbaud to determine the robustness of the
detector with respect to residual dispersion. The memoryless pdf loses
its accuracy for high symbol rates due to the strong interaction
between nonlinearities and dispersion and therefore the superiority of
the proposed detector disappears for these parameters and symbol rates
higher than 3 Gbaud. Similar observations regarding the accuracy of
the memoryless model have been made in \cite{beygi_JLT_DM_DP}. 
%

\section{Conclusion} 

\label{SEC:7} 

A low-complexity detector is proposed for memoryless
polarization-multiplexed fiber-optical channels by compensating the
amplitude-dependent NLPN. The compensation is performed by a phase
rotation of the received symbols depending on the amplitudes in both
polarizations. 
The performance results confirm the
superiority of PM schemes to SP for the same data rate. 




\end{document}